\pdfoutput=1

\documentclass[sigconf,nonacm]{acmart}

\acmYear{2019}
\startPage{1}

\setcopyright{none}

\usepackage{booktabs}   
\usepackage{subcaption} 

\usepackage{pgfplots}
\pgfplotsset{width=8.5cm,compat=1.8}

\usetikzlibrary{positioning,shadows.blur}

\usepackage{listings}
\usepackage{xcolor} 

\usepackage[linesnumbered,lined,boxed,commentsnumbered]{algorithm2e}

\begin{document}

\title[ClangJIT]{ClangJIT: Enhancing C++ with Just-in-Time Compilation} 


\author{Hal Finkel}
\orcid{0000-0002-7551-7122}             
\affiliation{
  \position{Lead, Compiler Technology and Programming Languages}
  \department{Leadership Computing Facility}              
  \institution{Argonne National Laboratory}            
  \streetaddress{9700 S Cass Ave}
  \city{Lemont}
  \state{IL}
  \postcode{60439}
  \country{USA}                    
}
\email{hfinkel@anl.gov}          

\author{David Poliakoff}
\affiliation{
  \institution{Lawrence Livermore National Laboratory}            
  \streetaddress{7000 East Avenue}
  \city{Livermore}
  \state{CA}
  \postcode{94550}
  \country{USA}                    
}
\email{poliakoff1@llnl.gov}          

\author{David F. Richards}
\affiliation{
  \institution{Lawrence Livermore National Laboratory}            
  \streetaddress{7000 East Avenue}
  \city{Livermore}
  \state{CA}
  \postcode{94550}
  \country{USA}                    
}
\email{richards12@llnl.gov}          


\begin{abstract}
The C++ programming language is not only a keystone of the high-performance-computing ecosystem but has proven to be a successful base for portable parallel-programming frameworks. As is well known, C++ programmers use templates to specialize algorithms, thus allowing the compiler to generate highly-efficient code for specific parameters, data structures, and so on. This capability has been limited to those specializations that can be identified when the application is compiled, and in many critical cases, compiling all potentially-relevant specializations is not practical. ClangJIT provides a well-integrated C++ language extension allowing template-based specialization to occur during program execution. This capability has been implemented for use in large-scale applications, and we demonstrate that just-in-time-compilation-based dynamic specialization can be integrated into applications, often requiring minimal changes (or no changes) to the applications themselves, providing significant performance improvements, programmer-productivity improvements, and decreased compilation time.
\end{abstract}



\maketitle

\begin{section}{Introduction and Related Work}
\label{sect:intro}

The C++ programming language is well-known for its design doctrine of leaving no room below it for another portable programming language~\cite{stroustrup2007evolving}. As compiler technology has advanced, however, it has become clear that using just-in-time (JIT) compilation to provide runtime specialization can provide performance benefits practically unachievable with purely ahead-of-time (AoT) compilation. Some of the benefits of runtime specialization can be realized using aggressive multiversioning, both manual and automatic, along with runtime dispatch across these pre-generated code variants (this technique has been applied for many decades; e.g., ~\cite{byler1987multiple}). This technique, however, comes with combinatorial compile-time cost, and as a result, is practically limited. Moreover, most of this compile-time cost from multiversioning is wasted when only a small subset of the specialized variants are actually used (as is often the case in practice). This paper introduces ClangJIT, an extension to the Clang C++ compiler which integrates just-in-time compilation into the otherwise-ahead-of-time-compiled C++ programming language. ClangJIT allows programmers to take advantage of their existing body of C++ code, but critically, defer the generation and optimization of template specializations until runtime using a relatively-natural extension to the core C++ programming language.

A significant design requirement for ClangJIT is that the runtime-compilation process not explicitly access the file system - only loading data from the running binary is permitted - which allows for deployment within environments where file-system access is either unavailable or prohibitively expensive. In addition, this requirement maintains the redistributibility of the binaries using the JIT-compilation features (i.e., they can run on systems where the source code is unavailable). For example, on large HPC deployments, especially on supercomputers with distributed file systems, extreme care is required whenever many simultaneously-running processes might access many files on the file system, and ClangJIT should elide these concerns altogether.

Moreover, as discussed below, ClangJIT achieves another important design goal of maximal, incremental reuse of the state of the compiler. As is well known, compiling C++ code, especially when many templates are involved, can be a slow process. ClangJIT does not start this process afresh for the entire translation unit each time a new template instantiation is requested. Instead, it starts with the state generated by the AoT compilation process, and from there, adds to it incrementally as new instantiations are required. This minimizes the time associated with the runtime compilation process.

ClangJIT is available online, see for more information: \\ \url{https://github.com/hfinkel/llvm-project-cxxjit/wiki}

\begin{subsection}{Related Work}
Clang, and thus ClangJIT, are built on top of the LLVM compiler infrastructure~\cite{lattner2004llvm}. The LLVM compiler infrastructure has been specifically designed to support both AoT and JIT compilation, and has been used to implement JIT compilers for a variety of languages both general purpose (e.g., Java~\cite{Falcon}, Haskell~\cite{terei2010llvm}, Lua~\cite{pall2008luajit}, Julia~\cite{bezanson2012julia}), and domain specific (e.g., TensorFlow/XLA~\cite{abadi2017computational}). In addition, C++ libraries implemented using LLVM to provide runtime specialization for specific domains are not uncommon (e.g., TensorFlow/XLA, Halide~\cite{ragan2013halide}).

Several existing projects have been developed to add dynamic capabilities to the C++ programming language~\cite{RuntimeCompiledCPlusPlusAlt}. A significant number of these rely on running the compiler as a separate process in order to generate a shared library, and that shared library is then dynamically loaded into the running process (e.g., ~\cite{binks2013runtime,noack2017kart}). Unfortunately, not only do such systems require being able to execute a compiler with access to the relevant source files at runtime, but careful management is required to constrain the cost of the runtime compilation processes, because each time the compiler is spawned it must start processing its inputs afresh. The Easy::JIT project~\cite{caamano2018easy} provides a limited JIT-based runtime specialization capability for C++ lambda functions to Clang, but this specialization is limited to parameter values, because the types are fixed during AoT compilation (a fork of Easy::JIT known as atJIT~\cite{atJIT} adds some autotuning capabilities). NativeJIT~\cite{NativeJIT} provides an in-process JIT for a subset of C for x86\_64. The closest known work to ClangJIT is CERN's Cling~\cite{vasilev2012cling} project. Cling also implements a JIT for C++ code using a modified version of Clang, but Cling's goals are very different from the goals of ClangJIT. Cling effectively turns C++ into a JIT-compiled scripting language, including a REPL interface. ClangJIT, on the other hand, is designed for high-performance, incremental compilation of template instantiations using only information embedded in the hosting binary. Cling provides a dynamic compiler for C++ code, while ClangJIT provides a language extension for embedded JIT compilation, and as a result, the two serve different use cases and have significantly-different implementations.

\end{subsection}

The hypothesis of this work is that, for production-relevant C++ libraries used for this kind of specialization, incremental JIT compilation can produce performance benefits while simultaneously decreasing compilation time. The initial evaluations presented in Section~\ref{sect:eval} confirm this hypothesis, and future work will explore applicability and benefits in more detail.

The remainder of this paper is structured as follows: Section~\ref{sect:lang} discusses the syntax and semantics of ClangJIT's language extension, Section~\ref{sect:aotpart} describes the implementation of ClangJIT's ahead-of-time-compilation components, Section~\ref{sect:rtpart} describes the implementation of ClangJIT's runtime components, Section~\ref{sect:eval} contains initial evaluation results, we discuss future work in Section~\ref{sect:fw}, and the paper concludes in Section~\ref{sect:conc}. 

\end{section}

\begin{section}{The Language Extension}
\label{sect:lang}

A key design goal for ClangJIT is natural integration with the C++ language while making JIT compilation easy to use. A user can enable JIT-compilation support in the compiler simply by using the command line flang {\tt -fjit}. Using this flag, both when compiling and when linking, is all that should be necessary to make using the JIT-compilation features possible. By itself, however, the command-line flag does not enable any use of JIT compilation. To do that, function templates can be tagged for JIT compilation by using the C++ attribute {\tt [[clang::jit]]}. An attributed function template provides for additional features and restrictions. These features are:

\begin{itemize}
\item Instantiations of this function template will not be constructed at compile time, but rather, calling a specialization of the template, or taking the address of a specialization of the template, will trigger the instantiation and compilation of the template at runtime.
\item Non-constant expressions may be provided for the non-type template parameters, and these values will be used at runtime to construct the type of the requested instantiation. See Listing~\ref{code:ex1} for a simple example.
\item Type arguments to the template can be provided as strings. If the argument is implicitly convertible to a {\tt const char *}, then that conversion is performed, and the result is used to identify the requested type. Otherwise, if an object is provided, and that object has a member function named {\tt c\_str()}, and the result of that function can be converted to a {\tt const char *}, then the call and conversion (if necessary) are performed in order to get a string used to identify the type. The string is parsed and analyzed to identify the type in the declaration context of the parent of the function triggering the instantiation. Whether types defined after the point in the source code that triggers the instantiation are available is not specified. See Listing~\ref{code:ex2} for a demonstration of this functionality, with Listing~\ref{code:ex2out} showing some example output.
\end{itemize}

\begin{lstlisting}[language=C++, caption={A JIT "hello world".}, label=code:ex1]
#include <iostream>
#include <cstdlib>

template <int x>
[[clang::jit]] void run() {
  std::cout << "Hello, World, I was compiled at runtime, x = " << x << "\n";
}

int main(int argc, char *argv[]) {
  int a = std::atoi(argv[1]);
  run<a>();
}
\end{lstlisting}

\begin{lstlisting}[language=C++, caption={A JIT example demonstrating using strings as types.}, label=code:ex2]
#include <iostream>

struct F {
  int i;
  double d;
};

template <typename T, int S>
struct G {
  T arr[S];
};

template <typename T>
[[clang::jit]] void run() {
  std::cout << "I was compiled at runtime, sizeof(T) = " << sizeof(T) << "\n";
}

int main(int argc, char *argv[]) {
  std::string t(argv[1]);
  run<t>();
}
\end{lstlisting}

\begin{lstlisting}[language=bash, caption={Output from the strings-as-types example in Listing~\ref{code:ex2}.}, label=code:ex2out]
$ clang++ -O3 -fjit -o /tmp/jit-t /tmp/jit-t.cpp
$ /tmp/jit-t '::F'
I was compiled at runtime, sizeof(T) = 16
$ /tmp/jit-t 'F'
I was compiled at runtime, sizeof(T) = 16
$ /tmp/jit-t 'float'
I was compiled at runtime, sizeof(T) = 4
$ /tmp/jit-t 'double'
I was compiled at runtime, sizeof(T) = 8
$ /tmp/jit-t 'size_t'
I was compiled at runtime, sizeof(T) = 8
$ /tmp/jit-t 'std::size_t'
I was compiled at runtime, sizeof(T) = 8
$ /tmp/jit-t 'G<F, 5>'
I was compiled at runtime, sizeof(T) = 80
\end{lstlisting}

There are a few noteworthy restrictions:
\begin{itemize}
\item Because the body of the template is not instantiated at compile time, {\tt decltype(auto)} and any other type-deduction mechanisms depending on the body of the function are not available.
\item Because the template specializations are not compiled until runtime, they're not available at compile time for use as non-type template arguments, etc.
\end{itemize}

Explicit specializations of a JIT function template are not JIT compiled, but rather, compiled during the regular AoT compilation process. If, at runtime, values are specified corresponding to some explicit specialization (which will have already been compiled), the template instantiation is not recompiled, but rather, the already-compiled function is used. An exception to this rule is that a JIT template with a pointer/reference non-type template parameter which is provided with a runtime pointer value will generate a different instantiation for each pointer value. If the pointer provided points to a global object, no attempt is made to map that pointer value back to the name of the global object when constructing the new type. This might seem like a bit of trivia, but has an important implication for the generated code. In general, pointer/reference-type non-type template arguments are not permitted to point to subobjects. This restriction still applies formally to the templates instantiated at runtime using runtime-provided pointer values. This has important optimization benefits: pointers that can be traced back to distinct underlying objects are known not to alias, and these template parameters appear to the optimizer to have this unique-object property. C++ does not yet have a {\tt restrict} feature, as C does, to represent the lack of pointer aliasing, but this aspect of combining JIT compilation with templates provides C++ with this feature in a limited way\footnote{Nearly all C++ compilers support some form of C's {\tt restrict} keyword as an extension, so this is not the {\it only} extension that provides this functionality.}.

\end{section}

\begin{section}{How it Works: Ahead-of-Time Compilation}
\label{sect:aotpart}

Implementing ClangJIT required modifying Clang's semantic-analysis and code-generation components in non-trivial ways. Clang's parsing and semantic analysis was extended to allow the {\tt [[clang::jit]]} attribute to appear on declarations and definitions of function templates. The most-significant modifications were to the code in Clang which determines whether a given template-argument list can be used to instantiate a given function template. In this case, when the function template in question has the {\tt [[clang::jit]]} attribute, two important changes were made: For a candidate non-type template argument (e.g., an expression of type int), the candidate is allowed to match without the usual check to determine if constant evaluation is possible. For a candidate template argument that should be a type, if the candidate is instead a non-type argument, logic was added to check for conversion to {\tt const char *}, first by calling a {\tt c\_str()} method if necessary, and if the conversion is possible, the candidate is allowed to match. Each time a JIT function template is instantiated, the instantiation is assigned a translation-unit-unique integer identifier which will be used during code generation and by the runtime library.

In order for the runtime library to instantiate templates dynamically, it requires a saved copy of Clang's abstract-syntax tree (AST), which is the internal data structure on which template instantiation is performed. Fortunately, Clang already contains the infrastructure for serializing and deserializing its AST, and moreover, embedding compressed copies of the input source files along with the AST, as part of the implementation of its modules feature. The embedded source files are important so that, should an error occur during template instantiation (e.g., a {\tt static\_assert} is triggered), useful messages can be produced for the user. Reconstructing the parameters used for code generation (e.g., whether use of AVX-2 vector instructions is enabled when targeting the x86\_64 architecture) also requires the set of command-line parameters passed by Clang's driver to the underlying compilation invocation. In addition, in order to allow JIT-compiled code to access local variables in what would otherwise be their containing translation unit, the addresses of such potentially-required variables are saved for use by the runtime library. All of these items are embedded in the compiled object file, resulting, as illustrated by Figure~\ref{fig:fatobj}, in a "fat" object file containing both the compiled host code as well as the the information necessary to resume the compilation process during program execution.

\begin{figure*}[h]
\caption{When using JIT-compilation features, the object files produced by the compiler become "fat" object files containing information needed at runtime.}
\label{fig:fatobj}
\centering
\begin{tikzpicture} 
   \tikzset{
     box/.style    = { rounded corners = 5pt,
                       align           = left,
                       font            = \sffamily\footnotesize,
                       text width      = 3.45cm, 
                       blur shadow     = {shadow blur steps = 15} },    
     legend/.style = { font       = \sffamily\bfseries, 
                       align      = right,
                       text width = 3.4cm},
  }
  \node [shade,
    blur shadow  = {shadow blur steps = 15},
    text width   = 1.01\textwidth,
    top color    = white, 
    bottom color = blue,
    text         = black, 
    font         = \sffamily\bfseries\large] (A)
    { \vspace{.3\textwidth}
     A JIT-enabled "fat" object file};
  
  \node [box, below left  = -5.25cm and -6.cm of A, fill = yellow, text width = 5.5cm]
    (DE)
    {\underline{\bfseries Serialized AST}
      \begin{itemize} 
        \setlength{\itemindent} {-.5cm}
        \item Preprocessor state and compressed source files
        \item Binary encoding of the AST
      \end{itemize}
    };

  \node [box, below = .25cm of DE, fill = yellow, text width = 5.5cm]
    (DE)
    {\underline{\bfseries Compilation Command-Line Arguments}
      \begin{itemize} 
        \setlength{\itemindent} {-.5cm}
        \item Used to restore code-generation options.
      \end{itemize}
    };

  \node [box, below = .25cm of DE, fill = yellow, text width = 5.5cm]
    (DE2)
    {\underline{\bfseries Optimized LLVM IR}
      \begin{itemize} 
        \setlength{\itemindent} {-.5cm}
        \item Used to allow inlining of pre-compiled functions into JIT-compiled functions.
      \end{itemize}
    };

  \node [box, below = .25cm of DE2, fill = yellow, text width = 5.5cm]
    (DE3)
    {\underline{\bfseries Local Symbol Addresses}
      \begin{itemize} 
        \setlength{\itemindent} {-.5cm}
        \item Used to allow the JIT to look up non-exported symbols in the translation unit.
      \end{itemize}
    };

  \node [box, below right  = -2cm and 1.5cm of DE, fill = yellow, text width = 6.25cm]
    (SD)
    { \underline{\bfseries Serialized AST for Device {\tiny (First Architecture)}}
    };

  \node [box, below  = 0.25cm of SD, fill = yellow, text width = 6.25cm]
    (SDC)
    { \underline{\bfseries Compilation Command-Line Arguments {\tiny (First Architecture)}}
    };

  \node [box, below  = 0.25cm of SDC, fill = yellow, text width = 6.25cm]
    (SDI)
    { \underline{\bfseries Compilation Optimized LLVM IR {\tiny (First Architecture)}}
    };

  \node [box, below right  = 0.25cm and -4.5cm of SDI, fill = yellow, text width = 6.5cm]
    (SD2)
    { \underline{\bfseries Serialized AST for Device {\tiny (Second Architecture)}} 
    };

  \node [box, below  = 0.25cm of SD2, fill = yellow, text width = 6.5cm]
    (SD2C)
    { \underline{\bfseries Compilation Command-Line Arguments {\tiny (Second Architecture)}}
    };

  \node [box, below  = 0.25cm of SD2C, fill = yellow, text width = 6.5cm]
    (SD2I)
    { \underline{\bfseries Compilation Optimized LLVM IR {\tiny (Second Architecture)}}
    };

  \node [below right = 0.5cm and -1.75cm of DE3, font = \sffamily\bfseries\large ] (d1) 
    {All Targets $\triangleleft$};

  \node [right = .5cm of d1, font = \sffamily\bfseries\large ] (d2) 
    {$\triangleright$ CUDA Support };
  
   \path [ draw, color = black, dashed, line width = 2pt ]
     (d1.south east) + (0.3cm,0)   coordinate(x1) -- (x1|-A.north);  
\end{tikzpicture}
\end{figure*}
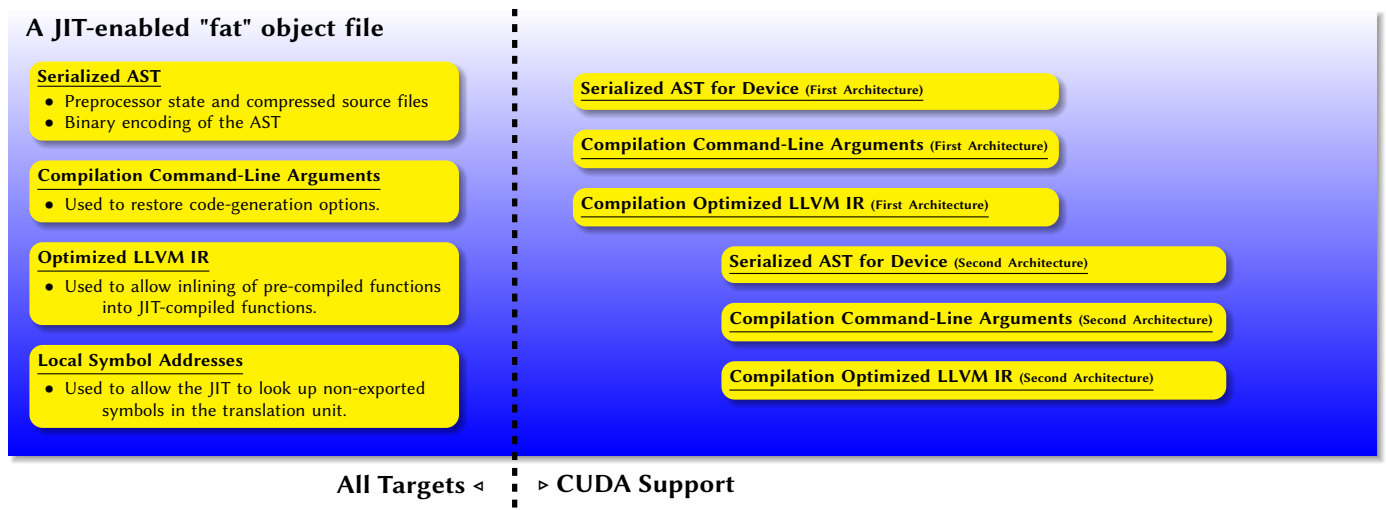

When emitting a call to a direct callee, and when getting a function pointer to a given function, Clang's code generation needs to generate a pointer to the relevant function. For JIT-compiled function-template instantiations, Clang is enhanced to generate a call to a special runtime function, {\tt \_\_clang\_jit}, which returns the required function pointer. How this runtime function works will be discussed in Section~\ref{sect:rtpart}. The runtime-dependent non-type template parameters for the particular template instantiation are packed into an on-stack structure, an on-stack array of strings representing runtime types is formed, and the addresses of these on-stack data structures are provided as arguments to the {\tt \_\_clang\_jit} call. Two additional parameters are passed: first, a mangled name for the template being instantiated (with wildcard character sequences in places where the runtime values and types are used), used in combination with the runtime values to identify the instantiation, and second, the translation-unit-unique identifier for this particular instantiation. This translation-unit-unique identifier is used by the runtime library to look up this particular instantiation in the serialized AST.

Finally, Clang's driver code was updated so that use of the {\tt -fjit} flag not only enables processing of the {\tt [[clang::jit]]} attributes during compilation, but also causes Clang's implementation libraries to be linked with the application, and when producing dynamically-linked executables, -rdynamic is implied. -rdynamic is used so that the runtime library can use the executable's exported dynamic symbol table to find external symbols from all translation units comprising the program (the alternative would require essentially duplicating this information in the array of local variables passed to the runtime library).

\begin{subsection}{CUDA Support}
In order to support JIT compilation of CUDA kernels for applications targeting NVIDIA GPUs, ClangJIT includes support for the CUDA programming model built on Clang's native CUDA support~\cite{wu2016gpucc}. The primary challenge in supporting CUDA in ClangJIT derives from the fact that, when compiling CUDA code, the driver invokes the compiler multiple times: Once to compile for the host architecture and once for each targeted GPU architecture (e.g., sm\_35). At runtime, the state of not only the host-targeting compiler invocation must be reconstructed, but also the state of one of these GPU-targeting compiler invocations (whichever most closely matches the device being used at runtime). Fortunately, Clang's CUDA compilation workflow compiles code for the GPUs first, and only once all GPU-targeting compilation is complete, is the compiler targeting the host invoked. When JIT compilation and CUDA are both enabled, the driver creates a temporary LLVM bitcode file, and each GPU-targeting compilation saves the serialized AST, command-line parameters, and some additional metadata into a set of global variables in this bitcode file. The implementation takes advantage of LLVM's {\it appending linkage} feature so that each GPU-targeting invocation can easily add pointers to its state data to an array of similar entries from all GPU-targeting invocations within the bitcode file. When the host-targeting compilation takes place, the bitcode file is loaded and linked into the LLVM module Clang is constructing and the address of the relevant global variable (which contains pointers to all of the other device-compilation-relevant global variables)  becomes an argument to calls to {\tt \_\_clang\_jit}. As illustrated by Figure~\ref{fig:fatobj}, all of this information ends up embedded in the host object file.

\end{subsection}

\end{section}

\begin{section}{How it Works: Runtime Compilation}
\label{sect:rtpart}

ClangJIT's runtime library is large, namely because it includes all of Clang and LLVM, but it has only one entry-point used by the compiler-generated code: {\tt \_\_clang\_jit}. This function is used to instantiate function templates at runtime. The ClangJIT-specific parts of the runtime library are approximately two thousand lines of code, and an outline of the implementation is provided in Algorithm~\ref{algo:cj}. The runtime library has a program-global cache of generated instantiations, but otherwise keeps separate state for each translation unit making use of JIT compilation. This per-translation-unit state is largely composed of two parts: first, a Clang compiler instance holding the AST and other data structures, and second, an in-memory LLVM IR module containing externally-available definitions. This LLVM IR module is initially populated by loading the optimized LLVM IR for the translation unit that is stored in the running binary, and marking all definitions with externally-available linkage, thus allowing the definitions to be analyzed and inlined into newly-generated code.

When the {\tt \_\_clang\_jit} function is called to retrieve a pointer to a requested template instantiation, the instantiation is looked up in the cache of instantiations. This cache uses LLVM's {\tt DenseMap} data structure, along with a mutex, and while these constructs are designed to have high performance, this lookup can have noticeable overhead. If the instantation does not exist, as outlined in Algorithm~\ref{algo:cj}, the instantiation is created along with any other new dependencies (e.g., a static function not otherwise used in the translation unit might now need to be emitted along with the requested instantiation). A new LLVM IR module is created by Clang, and that LLVM IR module is merged with the module containing the externally-available definitions. This combined module is optimized and JIT compiled. The newly-generated LLVM IR is then also merged, with externally-available linkage, into the module containing externally-available definitions for all previously-generated code. These externally-available definitions form a kind of cache used to enable inlining of previously-generated code into newly-generated code. This is important because if the incremental code generation associated with JIT compilation could not inline code generated in other stages, then the JIT-compiled code would likely be slower than the AoT-compiled code.

\IncMargin{1em}
\begin{algorithm}
\SetKwData{Left}{left}\SetKwData{This}{this}\SetKwData{Up}{up}
\SetKwFunction{Union}{Union}\SetKwFunction{FindCompress}{FindCompress}

\SetKwInOut{Input}{input}\SetKwInOut{Output}{output}\SetKwInOut{Globals}{globals}
\Input{TranslationUnitData\{SerializedAST, CommandLineArguments, OptimizedLLVMIR, LocalSymbolsArray\}, NonTypeTemplateValues, TypeStringsArray, MangledName, InstantiationIdentifier}
\Output{A pointer to the requested function}
\Globals{Instantiations, TUCompilerInstances}
\BlankLine

\lIf{\emph{Instantiations.lookup\{MangledName, NonTypeTemplateValues, TypeStringsArray\}}}{return the cached function pointer}

\lIf{\emph{A compiler instance for the translation unit of the caller $\not\in$ TUCompilerInstances}}{create a compiler instance using TranslationUnitData, loading the optimized LLVM IR and marking all definitions as {\it externally available}, and add the instance to TUCompilerInstances}

use InstantiationIdentifier to look up the instantiation, and then its parent template, in the AST

form a new template argument list using the instantiation in the AST combined with the values in NonTypeTemplateValues and the types in TypeStringsArray

call Clang's internal InstantiateFunctionDefinition function to get a new AST function definition node

\Repeat{no new functions were just emitted to LLVM IR}{
  emit all functions in Clang's deferred list which don't exist in the program to LLVM IR

  \For{all LLVM IR declarations for symbols that don't exist in the program}{
    lookup the AST node for the declaration and call Clang's internal HandleTopLevelDecl function on the AST node 
  }
}

call Clang's internal HandleTranslationUnit function to finalize the LLVM IR module being constructed

mark nearly everything in the new LLVM IR module as having {\it external} linkage

merge the {\it externally available} LLVM IR definitions into the just-generated LLVM IR module

optimize the new LLVM IR module and provide the result to LLVM's Orc JIT engine

take the newly-generated LLVM IR, mark the definitions as {\it externally available}, and merge with the module containing the other externally-available definitions

get the function pointer for the requested instantiaton, store it in Instantiations, and return it
\caption{{\tt \_\_clang\_jit} - Retrieve a JIT-Compiled Function-Template Instantiation}\label{algo:cj}
\end{algorithm}\DecMargin{1em}

Clang already uses LLVM's virtual-file-system infrastructure and this allows isolating it from any file-system access at runtime. As noted earler, it is important to avoid JIT compilation triggering any file-system access, and so, the Clang compiler instance created by the runtime library is provided only with an in-memory virtual-file-system provider. That provider only contains specific in-memory files with data from the running binary's compiler-created global variables.

\begin{subsection}{CUDA Support}
To support JIT compilation of CUDA code, which, in interesting cases, will require JIT compilation of CUDA kernels, dedicated logic exists in ClangJIT's runtime library. First, the AoT-compilation might have compiled device code for multiple device architectures (e.g., for sm\_35 and sm\_70). The CUDA runtime library is queried in order to determine the {\it compute capability} of the current device, and based on that, ClangJIT's runtime selects which device compiler state to use for JIT compilation. As for the host, the serialized AST, command-line options, and optimized LLVM IR are loaded to create a device-targeting compiler instance. During the execution of Algorithm~\ref{algo:cj}, at a high level, whatever is done with the host compiler instance is also done with the device compiler instance. However, instead of taking the optimized IR and handing it off to LLVM's JIT engine, the device compiler instance is configured to run the NVPTX backend and generate textual PTX code. This PTX code is then wrapped in an in-memory CUDA fatbin file\footnote{Unfortunately, the format of the fatbin files is not documented by NVIDIA, but given the information available in \url{https://reviews.llvm.org/D8397} and in~\cite{diamos2010ocelot}, we were able to create fatbins that are functional for this purpose.}, and that fatbin file is provided to the host compiler instance to embed in the generated module in the usual manner for CUDA compilation.
\end{subsection}

\end{section}

\begin{section}{Initial Evaluation}
\label{sect:eval}

We present here an evaluation of ClangJIT on three bases: ease of use, compile-time reduction, and runtime-performance improvement. First, we'll discuss the performance of ClangJIT by making use of a microbenchmark\footnote{The microbenchmark presented here is an adapted version of \url{https://github.com/eigenteam/eigen-git-mirror/blob/master/bench/benchmark.cpp}.} which relies on the Eigen C++ matrix library~\cite{guennebaud2010eigen}. Consider a simple benchmark which repeatedly calculates $M = I + 5\times 10^{-5}(M + M^2)$ for a matrix $M$ of some $NxN$ size. We're interested here in cases where $N$ is small, and examples from real applications where such small loop bounds occur will be presented in the following sections. Listing~\ref{code:eigen1} excerpts the version of the benchmark where dynamic matrix sizes are handled in the traditional manner. Listing~\ref{code:eigen2} excerpts the version of the benchmark where code for dynamic matrix sizes is generated at runtime using JIT compilation.

\begin{lstlisting}[language=C++, caption={An excerpt from the simple Eigen benchmark.}, label=code:eigen1]
#include <Eigen/Core>

using namespace std;
using namespace Eigen;

template <typename T>
void test_aot(int size, int repeat) {
  Matrix<T,Dynamic,Dynamic> I = Matrix<T,Dynamic,Dynamic>::Ones(size, size);
  Matrix<T,Dynamic,Dynamic> m(size, size);
  for(int i = 0; i < size; i++)
  for(int j = 0; j < size; j++) {
    m(i,j) = (i+size*j);
  }

  for (int r = 0; r < repeat; ++r) {
    m = Matrix<T,Dynamic,Dynamic>::Ones(size, size) + T(0.00005) * (m + (m*m));
  }
}

void test_aot(std::string &type, int size, int repeat) {
  if (type == "float")
    test_aot<float>(size, repeat);
  else if (type == "double")
    test_aot<double>(size, repeat);
  else if (type == "long double")
    test_aot<long double>(size, repeat);
  else
    cout << type << "not supported for AoT\n";
}
\end{lstlisting}

\begin{lstlisting}[language=C++, caption={An excerpt from the simple Eigen benchmark (JIT version).}, label=code:eigen2]
template <typename T, int size>
[[clang::jit]] void test_jit_sz(int repeat) {
  Matrix<T,size,size> I = Matrix<T,size,size>::Ones();
  Matrix<T,size,size> m;
  for(int i = 0; i < size; i++)
  for(int j = 0; j < size; j++) {
    m(i,j) = (i+size*j);
  }

  for (int r = 0; r < repeat; ++r) {
    m = Matrix<T,size,size>::Ones() + T(0.00005) * (m + (m*m));
  }
}

void test_jit(std::string &type, int size, int repeat) {
  return test_jit_sz<type, size>(repeat);
}
\end{lstlisting}

The Eigen library was chosen for this benchmark because the library supports matrices of both compile-time size (specified as non-type template parameters) and runtime size (specific as constructor arguments). When using JIT compilation, we can use the non-type-template-parameter method to specify sizes known only during program execution. First, we'll examine the compile-time advantages that ClangJIT offers over both traditional AoT compilation and over the up-front compilation of numerous potentially-used template specializations. In Figure~\ref{fig:ect}, we present the AoT compilation time for this benchmark in various configurations\footnote{This benchmarking was conducted on a Intel Xeon E5-2699 using the flags -march=native -ffast-math -O3, and using a ClangJIT build compiled using GCC 8.2.0 with CMake's RelWithDebInfo mode}. For all of these times, we subtracted a baseline compilation time of 2.58s - the time required to compile a trivial {\tt main()} function with the Eigen header file included. The time identified by "J" indicates the AoT compilation time for the benchmark when only the JIT-compilation-based implementation is present (i.e., that in Listing~\ref{code:eigen2} plus the associated main function). The time identified by "A1" indicates the compilation time for the generic version, compiled only using the scalar type {\tt double} (i.e., Listing~\ref{code:eigen1} but with only the {\tt double} case present). As can be seen, this takes significantly longer than the AoT compilation time for the JIT-based version. Moreover, the JIT-based version can be used with any scalar type. If we try to replicate that capability with the traditional AoT approach, and thus instantiate the implementation for {\tt float}, {\tt double}, and {\tt long double}, as shown in Listing~\ref{code:eigen1}, then the AoT compilation time is nearly 7x larger than using the JIT capability. The reported "J" time does omit the time spent at runtime compiling the necessary specializations. Here we show the compilation time for different specializations (i.e., with both the size and type fixed), for a scalar type of {\tt double}, where the size was 16 for "S16", the size was 7 for "S7", the size was 3 for "S3", and size was 1 for "S1", and specializations for both sizes 16 and 7 were included in the time "S16a7" (to demonstrate that the specialization compilation times are roughly additive). It is useful to note that the specialization compilation time depends on the size, but is always less than the single size-generic implementation in "A1". As the size becomes larger, the difference in the work the compiler must do to handle the specialization compared to handling the generic version shrinks.

\begin{figure}[h]
\caption{AoT compilation time of the Eigen benchmark. See the text for a description of the configurations presented.}
\label{fig:ect}
\centering
\begin{tikzpicture}
\begin{axis}[
    ybar,
    enlargelimits=0.15,
    legend style={at={(0.5,-0.15)},
      anchor=north,legend columns=-1},
    ylabel={time (s)},
    symbolic x coords={J,S16,S7,S3,S1,S16a7,A3,A1},
    xtick=data,
    nodes near coords,
    nodes near coords align={vertical},
    every node near coord/.append style={font=\tiny},
    ]
\addplot coordinates {(J,0.92) (S16,2.37) (S7,0.72) (S3,0.62) (S1,0.37) (S16a7,3.12) (A3,7.12) (A1,2.72)};
\end{axis}
\end{tikzpicture}
\end{figure}
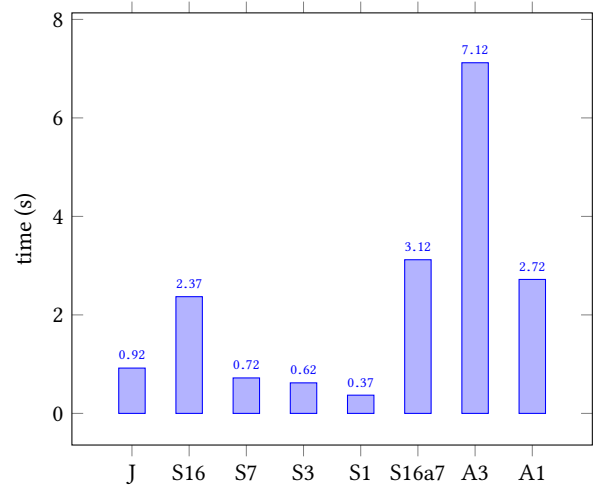

Figure~\ref{fig:eperf} shows the performance of the benchmark, using type {\tt double}, for several sizes. For small sizes, when the size is known, the compiler can unroll the loops and perform other optimizations to produce code that performs significantly better than the generic version. This performance is essentially identical to the performance of the AoT-generated specializations of the same size, but of course, the JIT-based version is more flexible. As the size gets larger, the code the compiler generates becomes increasingly similar to that generated for the generic version, and differences such as generated tail loops come to have a decreasingly-important performance impact, and so for large sizes little performance difference remains between the JIT-compiled code and the generic version.

\begin{figure}[h]
\caption{Runtime performance of the Eigen benchmark, using type {\tt double}, for the indicated size. The time is normalized to the time of the JIT-compiled version. For small sizes, the JIT-compiled version is significantly faster.}
\label{fig:eperf}
\centering
\begin{tikzpicture}
\begin{axis}[
    ybar,
    enlargelimits=0.15,
    legend style={at={(0.5,-0.15)},
      anchor=north,legend columns=-1},
    ylabel={time relative to JIT},
    symbolic x coords={Sz3, Sz7, Sz16},
    xtick=data,
    nodes near coords,
    nodes near coords align={vertical},
    every node near coord/.append style={font=\tiny},
    ]

\addplot coordinates {(Sz3,1) (Sz7,1) (Sz16,1)};
\addplot coordinates {(Sz3,1.01) (Sz7,1.01) (Sz16,1)};
\addplot coordinates {(Sz3,8.05) (Sz7,2.4) (Sz16,1.02)};
\legend{JIT-compiled,AoT specialization,AoT generic}
\end{axis}
\end{tikzpicture}
\end{figure}
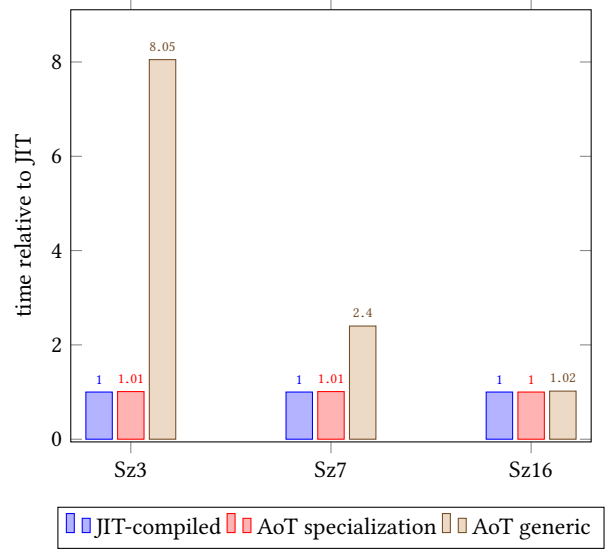

Figure~\ref{fig:ecudaperf} shows the performance of the benchmark, using type {\tt double}, adapted to use CUDA\footnote{The CUDA benchmark was run on an IBM POWER8 host with an NVIDIA Tesla K80 GPU using CUDA 9.2.148.}. The source code for the CUDA adaptation is straightforwardly derived from the original where the the matrix computation is moved into a kernel, and that kernel is executed using one GPU thread\footnote{A call to {\tt cudaThreadSetLimit(cudaLimitMallocHeapSize, $\ldots$)} was inserted in the non-JIT implementation to allow dynamic allocation to work on the device.}. This is meant to serve as a proxy for part of a larger computation, presumably running on many threads, and so we look at two metrics: First, we look at the runtime performance of the JIT-compiled kernel compared to the generic AoT-compiled kernel, and second, we look at the number of registers used by the various kernels. Using a large number of registers can limit GPU occupancy, and so when considered in the context of a larger calculation, if the JIT-compiled kernel uses a smaller number of registers than the AoT-compiled generic implementation, that adds additional performance benefits. As shown in Figure~\ref{fig:ecudaperf}, for matrix sizes $1x1$ through $7x7$, the serial performance of the JIT-compiled GPU kernels is one to two orders of magnitude better than the generic AoT-compiled version\footnote{The compilation of the Eigen matrix type for sizes larger than $7x7$ failed, because some required template specializations were not available, a problem that did not occur when compiling for the host, and so $7x7$ is the largest size shown.}. As shown in Figure~\ref{fig:ecudaregs}, on top of that, the smaller JIT-compiled kernels used far fewer registers than the generic AoT-compiled version\footnote{Repeating the experiment with type {\tt float} shows the JIT-compiled kernels always use fewer registers than the AoT-compiled generic version. Loop unrolling and other compiler optimizations can increase register pressure, so while the specialized code can use many fewer registers, specialization is not guaranteed to reduce register usage.}.

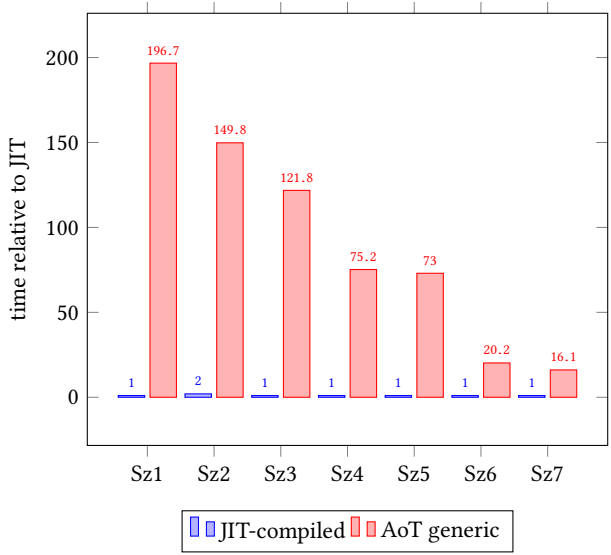
\begin{figure}[h]
\caption{Runtime performance of the Eigen benchmark using CUDA, using type {\tt double}, for the indicated size. The time is normalized to the time of the JIT-compiled version. For small sizes, the JIT-compiled version is significantly faster.}
\label{fig:ecudaperf}
\centering
\begin{tikzpicture}
\begin{axis}[
    ybar,
    enlargelimits=0.15,
    legend style={at={(0.5,-0.15)},
      anchor=north,legend columns=-1},
    ylabel={time relative to JIT},
    symbolic x coords={Sz1, Sz2, Sz3, Sz4, Sz5, Sz6, Sz7},
    xtick=data,
    nodes near coords,
    nodes near coords align={vertical},
    every node near coord/.append style={font=\tiny},
    ]

\addplot coordinates {(Sz1,1) (Sz2,2) (Sz3,1) (Sz4,1) (Sz5,1) (Sz6,1) (Sz7,1)};
\addplot coordinates {(Sz1,196.7) (Sz2,149.8) (Sz3,121.8) (Sz4,75.2) (Sz5,73) (Sz6,20.2) (Sz7,16.1)};
\legend{JIT-compiled,AoT generic}
\end{axis}
\end{tikzpicture}
\end{figure}

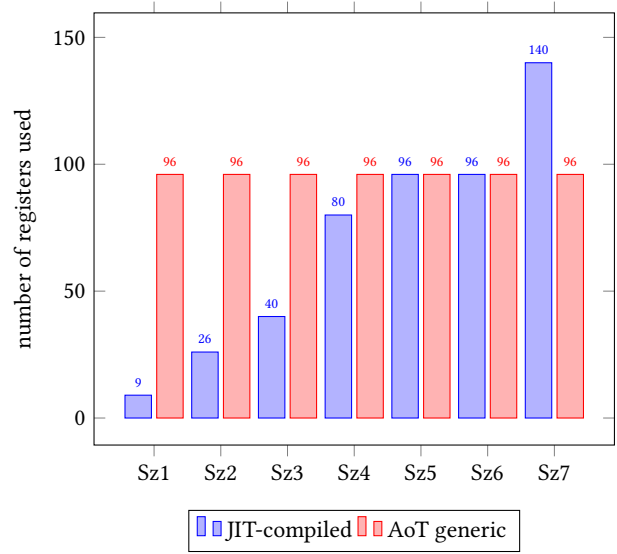
\begin{figure}[h]
\caption{Register usage of the Eigen benchmark using CUDA, using type {\tt double}, for the indicated size.}
\label{fig:ecudaregs}
\centering
\begin{tikzpicture}
\begin{axis}[
    ybar,
    enlargelimits=0.15,
    legend style={at={(0.5,-0.15)},
      anchor=north,legend columns=-1},
    ylabel={number of registers used},
    symbolic x coords={Sz1, Sz2, Sz3, Sz4, Sz5, Sz6, Sz7},
    xtick=data,
    nodes near coords,
    nodes near coords align={vertical},
    every node near coord/.append style={font=\tiny},
    ]

\addplot coordinates {(Sz1,9) (Sz2,26) (Sz3,40) (Sz4,80) (Sz5,96) (Sz6,96) (Sz7,140)};
\addplot coordinates {(Sz1,96) (Sz2,96) (Sz3,96) (Sz4,96) (Sz5,96) (Sz6,96) (Sz7,96)};
\legend{JIT-compiled,AoT generic}
\end{axis}
\end{tikzpicture}
\end{figure}

In the following we present a case study on using JIT compilation transparently, and then two case studies where ClangJIT was used to improve two open-source proxy applications developed by Lawrence Livermore National Laboratory (LLNL): Kripke and Laghos. These proxy applications make use of the RAJA library~\cite{hornung2014raja}, also developed at LLNL, as do the corresponding production applications. Care was taken to ensure that the techniques used to adapt these proxy applications to make use of ClangJIT can be applied to the production applications that they represent. Both compile-time and runtime performance data was collected with the assistance of the Caliper~\cite{boehme2016caliper} tool.

\begin{subsection}{RAJA and Transparent Usage}
The RAJA library aims to abstract underlying parallel programming models, such as OpenMP and CUDA, allowing portable applications to be created by making use of RAJA's programming-model-neutral multi-dimensional-loop templates. It is also possible to hide the use of ClangJIT's JIT-compilation behind RAJA's abstraction layer, allowing applications to make use of JIT compilation without changes to the application's source code. To illustrate how this works, consider the simple loop abstraction shown in Listing~\ref{code:simploop}.

\begin{lstlisting}[language=C++, caption={A simple loop abstraction.}, label=code:simploop]
  template<typename Body>
  void forall(int begin, int end, Body body) {
    for(int x = begin; x<end; x++) {
      body(x);
    }
  }

  // This template can be used like this:
  forall(begin_index, end_index, [=](int iter) {
    // Some physics goes here.
  });
\end{lstlisting}

A user might desire that the loop is compiled at runtime, allowing the optimizer to specialize the code for the particular index range provided. As shown in Listing~\ref{code:simpjitloop}, this can be done by wrapping the template using one marked for JIT compilation, and in doing so, allows using JIT compilation without changing the interface, the {\tt forall} template in this illustration, used by the application.

\begin{lstlisting}[language=C++, caption={A simple loop abstraction transparently using JIT compilation.}, label=code:simpjitloop]
  template<typename Body>
  void forall_original(int begin, int end, Body body) {
    for(int x = begin; x<end; x++){
      body(x);
    }
  }

  template<int Begin, int End, typename Body>
  [[clang::jit]] void forall_shim(Body body) {
    forall_original(begin, end, body);
  }

  template<typename Body>
  void forall(int begin, int end, Body body) {
    forall_shim<begin,end>(body);
  }
\end{lstlisting}

This serves to illustrate how the JIT capability can be transparently used by an application using a RAJA-like abstraction. Doing this, however, exposes a number of tradeoffs. First, the JIT compilation itself takes time. Second, the process of looking up already-compiled template instantiations also has an overhead that can be significant compared to a compiled function doing very little computational work per invocation. In essentially all of the evaluations presented in this paper, it is this lookup overhead that is most important. To illustrate the interplay between these overheads and how a realistic RAJA abstraction around JIT compilation can be constructed, we present a simple matrix-multiplication benchmark in Listing~\ref{code:smallmm} which uses the RAJA library abstraction in Listing~\ref{code:rajaabs}. As can be seen, this abstraction is more complicated than that in Listing~\ref{code:simpjitloop}, because it uses parameter packs to handle RAJA multi-dimensional ranges. This abstraction code would be placed in the RAJA library, however, thus absolving the application developer from dealing with these complexities.

\begin{lstlisting}[language=C++, caption={A simple RAJA matrix-multiplication benchmark.}, label=code:smallmm]
using namespace RAJA::statement;
using RAJA::seq_exec;
using MatMulPolicy = RAJA::KernelPolicy<
  For<
    0,seq_exec,
    For<
      1,seq_exec,
      For<
        2, seq_exec,
          For<3, seq_exec,Lambda<0>>
      >
    >
  >
>;

#define MAT2D(r,c,size) r*size+c

int main(int argc, char* argv[]) {
  std::size_t size = ...;
  std::size_t batch_size = ...;
  std::size_t repeats = ...;
  ... // Allocate and randomly initialize out_matrix, input_matrix1, input_matrix2
  for(long rep = 0; rep<(repeats/batch_size); rep++){
    affine_jit_kernel_difficult<MatMulPolicy>(
        camp::make_tuple(
          RAJA::RangeSegment(0,batch_size),
          RAJA::RangeSegment(0,size),
          RAJA::RangeSegment(0,size),
          RAJA::RangeSegment(0,size)
        ),
        [=](const std::size_t matrices, const std::size_t i, const std::size_t j,
            const std::size_t k){
            out_matrix[MAT2D(i,j,size)] +=
              input_matrix1[MAT2D(i,k,size)] * input_matrix2[MAT2D(k,j,size)];
        }
    );
}
\end{lstlisting}

\begin{lstlisting}[language=C++, caption={A RAJA abstraction to apply JIT compilation to a multi-dimensional loop kernel.}, label=code:rajaabs]
template<typename Policy, long... ends, typename Kernel>
[[clang::jit]] void affine_jit_kernel_full(Kernel&& kernel) {
  static auto rs =
    camp::make_tuple(
        RAJA::RangeSegment(0,ends)...
    );
  RAJA::kernel<Policy>(
      rs,
      std::forward<Kernel>(kernel)
  );
}

template<typename Policy, typename Kernel, typename... Args>
void affine_jit_kernel_difficult_helper2(Kernel&& kernel, Args... args) {
   affine_jit_kernel_full<Policy,*std::end(args)...>(
       std::forward<Kernel>(kernel)
   );
}

template<typename Policy, typename TupleLike, typename Kernel,
         std::size_t... indices>
void affine_jit_kernel_difficult_helper(TupleLike IndexTuple, Kernel&& kernel,
                                        std::index_sequence<indices...>) {
  affine_jit_kernel_difficult_helper2<Policy>(
   std::forward<Kernel>(kernel),
   camp::get<indices>(IndexTuple)...
  );
    
}

template<typename Policy, typename TupleLike, typename Kernel>
void affine_jit_kernel_difficult(TupleLike IndexTuple, Kernel&& kernel) {
  affine_jit_kernel_difficult_helper<Policy,TupleLike>(
      std::forward<TupleLike>(IndexTuple),
      std::forward<Kernel>(kernel),
      std::make_index_sequence<camp::tuple_size<TupleLike>::value>()
  );
}
\end{lstlisting}

We present in Figure~\ref{fig:mmperf} performance data from around the transition to profitability for various small sizes, for various batch sizes (which change based upon over how many invocations the lookup of the already-compiled template instantiation is amortized), and for different numbers of total iterations (which change based upon over how much work the JIT compilation itself is amortized). As can be seen, the compilation time can be significant if not enough computational work is performed by the compiled code (as is the case by the $size=2x2$ runs with fewer than $10^9$ total iterations), and moreover, the instantiation lookup adds significant overhead (as can be seen by noting that the speedup of the runs with smaller batch sizes is generally smaller than those with the larger batch sizes). 

\begin{figure}[h]
\caption{Performance of the RAJA small-matrix-multiply benchmark. The runs are labeled as b{\tt{}M}i{\tt{}T} where the batch size is $10^M$, and the total number of iterations is $10^T$.}
\label{fig:mmperf}
\centering
\begin{tikzpicture}
\begin{axis}[
    ybar,
    enlargelimits=0.15,
    legend style={at={(0.5,-0.15)},
      anchor=north,legend columns=-1},
    ylabel={relative speedup},
    symbolic x coords={b1i8,b3i8,b4i8,b3i9,b4i9},
    xtick=data,
    nodes near coords,
    nodes near coords align={vertical},
    every node near coord/.append style={font=\tiny},
    ]
\addplot coordinates {(b1i8,0.426858) (b3i8,0.827628) (b4i8,0.831693) (b3i9,1.077180) (b4i9,1.140284)};
\addplot coordinates {(b1i8,1.149796) (b3i8,1.172624) (b4i8,1.221582) (b3i9,1.219257) (b4i9,1.217480)};
\legend{size=2x2,size=8x8}
\end{axis}
\end{tikzpicture}
\end{figure}
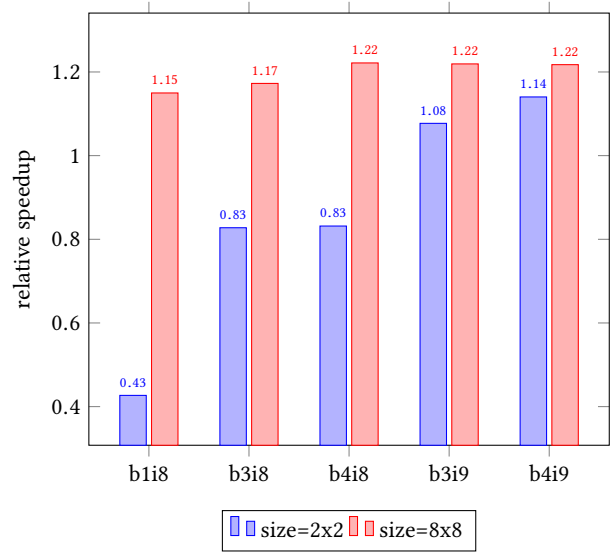

\end{subsection}

\begin{subsection}{Application: Kripke}
Kripke~\cite{kunen2015kripke} is a proxy application representing three-dimensional deterministic transport applications. Such an application has many nested loops which iterate over {\it directions}, {\it zones}, and {\it energy groups}. On different architectures, the optimal nesting order of those loops changes, as do the layouts of many data structures within the application. Managing this in an abstract way pushes Kripke to use bleeding-edge features from RAJA and modern C++. Unfortunately, constraints of C++ make such abstractions unfriendly to write, and as shown in Listing~\ref{code:kripold}, Kripke has to maintain a lot of plumbing code to select among all the possible layouts and loop-execution mechanisms based on runtime user selections. This also means that every possible variant of the loop has to be compiled ahead of time.

With a JIT compiler, however, runtime user selections are trivial, and as illustrated in Listing~\ref{code:kripnew}, these selections can be replaced with a much simpler dispatch mechanism, and the compilation of these selections happens only as needed. Moreover, removing the extra kernel variants speeds up the AoT compilation of many files by 2-3x\footnote{Specifically, the files Scattering.cpp (68\% compilation-time decrease), LTimes.cpp (58\% compilation-time decrease), and LPlusTimes.cpp (56\% compilation-time decrease), from which specializations were removed and replaced by uses of JIT compilation, exhibited AoT-compilation-time improvements.}, and leads to code that is easier to write and maintain.

\begin{lstlisting}[language=C++, caption={An excerpt from Kripke's ArchLayout.h and SteadyStateSolver.cpp, showing the original dispatching scheme.}, label=code:kripold]
  template<typename Function, typename ... Args>
  void dispatchLayout(LayoutV layout_v, Function const &fcn, Args &&... args)
  {
    switch(layout_v){
      case LayoutV_DGZ: fcn(LayoutT_DGZ{}, std::forward<Args>(args)...); break;
      case LayoutV_DZG: fcn(LayoutT_DZG{}, std::forward<Args>(args)...); break;
      ... // There are six lines, in total, like those above.
      default: KRIPKE_ABORT("Unknown layout_v=%d\n", (int)layout_v); break;
    }
  }

  template<typename Function, typename ... Args>
  void dispatchArch(ArchV arch_v, Function const &fcn, Args &&... args)
  {
    switch(arch_v){
      case ArchV_Sequential: fcn(ArchT_Sequential{}, std::forward<Args>(args)...);
        break;
#ifdef KRIPKE_USE_OPENMP
      case ArchV_OpenMP: fcn(ArchT_OpenMP{}, std::forward<Args>(args)...); break;
#endif 
      ... // A similar CUDA option.
      default: KRIPKE_ABORT("Unknown arch_v=%d\n", (int)arch_v); break;
    }
  }

  template<typename arch_t>
  struct DispatchHelper{
    template<typename layout_t, typename Function, typename ... Args>
    void operator()(layout_t, Function const &fcn, Args &&... args) const {
      using al_t = ArchLayoutT<arch_t, layout_t>;
      fcn(al_t{}, std::forward<Args>(args)...);
    }
  };

  template<typename Function, typename ... Args>
  void dispatch(ArchLayoutV al_v, Function const &fcn, Args &&... args)
  {
    dispatchArch(al_v.arch_v, [&](auto arch_t){
      DispatchHelper<decltype(arch_t)> helper;
      dispatchLayout(al_v.layout_v, helper, fcn, std::forward<Args>(args)...);
  });

  // Code like this appears several times to launch the specialized kernels.
  Kripke::dispatch(al_v, SourceSdom{}, sdom_id,
                     ... // Other parameters omitted.
                     source_strength);
}
\end{lstlisting}

\begin{lstlisting}[language=C++, caption={An excerpt from Kripke's SteadyStateSolver.cpp, adapted to use ClangJIT, showing how simple the dispatcher has become.}, label=code:kripnew]
  template<typename Body>
  void launcher(Body && body){
    body(std::string("Kripke::ArchLayoutT<Kripke::ArchT_Sequential,"
                     "Kripke::LayoutT_DGZ>"));
    // In the real application, a configuration file provides this string.
  }

  ...

  // Code like this now appears several times to lauch various JIT-compiled kernels.
  launcher([&](const std::string& ArchitectureLayout){
    Kripke::Kernel::template scattering<ArchitectureLayout>(data_store);
    // Notice that the ArchitectureLayout template parameter is a string.
  });
\end{lstlisting}

\end{subsection}

\begin{subsection}{Application: Laghos}
Laghos~\cite{laghos} is a higher-order finite-element-method (FEM) proxy application solving the time-dependent Euler equations of compressible gas dynamics in a moving Lagrangian frame using unstructured high-order finite-element spatial discretization~\cite{dobrev2012high} and explicit high-order time-stepping. Higher-order FEM codes are unlike traditional codes solving partial sdifferential equations in that, in addition to a traditional loop over all of the {\it elements} in a simulation, there are nested loops within each element whose bounds are determined at runtimes and often small. The relevant parameters (NUM\_DOFS\_1D and NUM\_QUAD\_1D in Listing~\ref{code:lagold}) are often between two and 32, and they form the bounds on over thirty loops in just one kernel, often to depths of four within the main element loop. Knowing these loop bounds is critical, as shown in Figure~\ref{fig:femperf}: not knowing them can cause an 8x slowdown.

Currently, as shown in Listing~\ref{code:lagold}, this performance opportunity is realized by picking the most common orders, instantiating a template function for each, and then dispatching among these explicitly-specialized functions. Since these codes already use template parameters, porting them to use ClangJIT was trivial. We match the performance of the template solution, which is 8x faster than the non-template solution, and make AoT compilation 10x faster. In our experience, higher-order finite-element codes use runtime specialization very frequently, and we believe that JIT compilation will play a huge part in doing this in a more-productive way in the future.

\begin{lstlisting}[language=C++, caption={The manual explicit-instantiation-and-dispatch code in Laghos's rMassMultAdd.cpp that ClangJIT makes obsolete.}, label=code:lagold]
  const unsigned int id = (DIM<<16)|((NUM_DOFS_1D-1)<<8)|(NUM_QUAD_1D>>1);
  static std::unordered_map<unsigned int, fMassMultAdd> call =
  {
    {0x20001,&rMassMultAdd2D<1,2>},    {0x20101,&rMassMultAdd2D<2,2>},
    {0x20102,&rMassMultAdd2D<2,4>},    {0x20202,&rMassMultAdd2D<3,4>},
    {0x20203,&rMassMultAdd2D<3,6>},    {0x20303,&rMassMultAdd2D<4,6>},
    {0x20304,&rMassMultAdd2D<4,8>},    {0x20404,&rMassMultAdd2D<5,8>},
    ... // There are approximately 32 lines, in total, like those above.
  };

  assert(call[id]); // This solution is not as flexible as using ClangJIT.
  call[id](numElements,dofToQuad,dofToQuadD,quadToDof,quadToDofD,op,x,y);
\end{lstlisting}

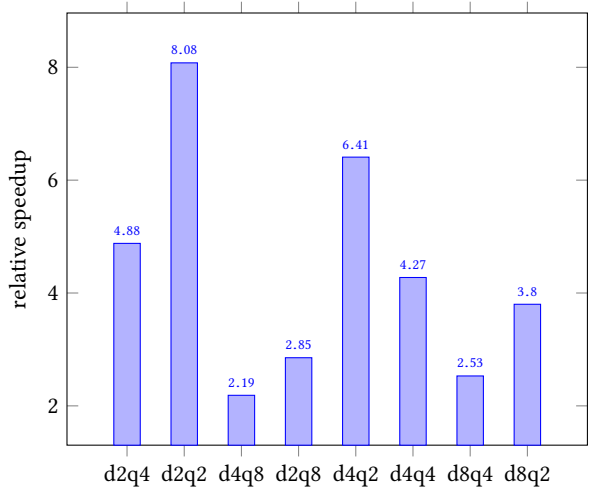
\begin{figure}[h]
\caption{Laghos runtime performance improvement using ClangJIT over the non-specialized kernel implementations, using 10000 elements. The runs are labled as d{\tt{}N}q{\tt{}M} where the {\tt dofs} parameter is {\tt N} and the {\tt quads} parameter is {\tt M}.}
\label{fig:femperf}
\centering
\begin{tikzpicture}
\begin{axis}[
    ybar,
    enlargelimits=0.15,
    legend style={at={(0.5,-0.15)},
      anchor=north,legend columns=-1},
    ylabel={relative speedup},
    symbolic x coords={d2q4,d2q2,d4q8,d2q8,d4q2,d4q4,d8q4,d8q2},
    xtick=data,
    nodes near coords,
    nodes near coords align={vertical},
    every node near coord/.append style={font=\tiny},
    ]
\addplot coordinates {(d2q4,4.879432) (d2q2,8.078300) (d4q8,2.187222) (d2q8,2.853074) (d4q2,6.407826) (d4q4,4.274835) (d8q4,2.530043) (d8q2,3.800424)};
\end{axis}
\end{tikzpicture}
\end{figure}

\end{subsection}

\end{section}

\begin{section}{Future Work}
\label{sect:fw}

Future enhancements to ClangJIT will allow for not just runtime specialization, but \emph{adaptive} runtime specialization. LLVM's optimizer can make good use of profiling information to guide function inlining, code placement, and other optimizations. ClangJIT can be enhanced to gather profiling information and use that information to recompile further-optimized code variants. LLVM supports two kinds of profiling: instrumentation-based profiling and sampling-based profiling, and how to best use this capability is under investigation. Moreover, profiling information can be used to guide more-advanced autotuning of the JIT-compiled code, and this is also under investigation.

LLVM contains a number of features that, while not used in support of traditional programming languages (e.g., C, C++, and Fortran) because of the corresponding runtime support required, are used to support dynamic languages, and could be applied to code compiled by ClangJIT at runtime. Implicit null checks, for example, which allow compiled code to elide nearly-always-false null-pointer checks by having the runtime library appropriately handle signals raised by rare memory-access violations, is a good candidate for inclusion in ClangJIT. Applying implicit null checks is dependent on having access to profiling information showing that the relevant checks essentially always succeed because they make the non-null case faster by making the null case very slow. Other LLVM features, such as those supporting dynamic patching and deoptimization, might also eventually be used with ClangJIT.

We will investigate enhancing ClangJIT with support for other accelerator programming models, especially HIP and OpenMP. HIP is a programming model for AMD GPUs, and its implementation in Clang is based on Clang's CUDA implementation. Because of the similarities between HIP and CUDA, we anticipate the changes necessary to add HIP support will be minor. OpenMP accelerator support, however, is more complicated. Even when targeting NVIDIA GPUs, Clang's OpenMP target-offload compilation pipeline differs significantly from its CUDA compilation pipeline. Among other differences, for OpenMP target offloading, the host code is compiled first, followed by the device code, and a linker script is used to integrate the various components later. Further enhancements to the scheme described in Section~\ref{sect:aotpart} are necessary in order to support a model where host compilation precedes device compilation.

During our explorations of different application use cases, two noteworthy enhancements were identified. First, C++ non-type template parameters currently cannot be floating-point values. However, for many HPC use cases, it will be useful to create specializations given specific sets of floating-point values (e.g., for some polynomial coefficients or constant matrix elements). Currently, this can be done by casting the floating-point values to integers, using those to instantiate the templates, and then casting the integers back to floating-point values inside the JIT-compiled code. The casting, however, is a workaround that we'll investigate eliminating. Second, the current implementation does not allow JIT-compiled templates to make use of other JIT-compiled templates - in other words, once ClangJIT starts compiling code at runtime, it assumes that code will not, itself, create more places where JIT compilation might be invoked in the future. This has limited applicability to some use cases because it creates an unhelpful barrier between code that can be used during AoT compilation and code that can be used (transitively) in JIT-compiled templates.

This work has inspired a recent proposal to the C++ standards committee~\cite{P1609R0}, and as discussed in that proposal, the language extension discussed here might not be the best way to design such an extension to C++. As highlighted in that proposal, use of an attribute such as {\tt [[clang::jit]]} might be suboptimal because:
\begin{itemize}
\item The template is not otherwise special, it is the point of instantiation that is special (and the point of instantiation is what might cause compilation to fail without JIT-compilation support).
\item The uses of the template look vaguely normal, and so places where the application might invoke the JIT compiler will be difficult to spot during code review.
\item The current mechanism provides no place to get out an error or provide a fall-back execution path - except that having the runtime throw an exception might work.
\end{itemize}
Future work will explore different ways to address these potential downsides of the language extension presented here. For example, it might be better to use some syntax like:
\begin{lstlisting}[language=C++,numbers=none,frame=none]
  jit_this_template foo<argc>();
\end{lstlisting}
where {\tt jit\_this\_template} would be a new keyword.

\end{section}

\begin{section}{Conclusions}
\label{sect:conc}
In this paper, we've demonstrated that JIT-compilation technology can be integrated into the C++ programming language, that this can be done in a way which makes using JIT compilation easy, that this can reduce compile time, making application developers more productive, and that this can be used to realize high performance in HPC applications. We investigated whether JIT compilation could be integrated into applications making use of the RAJA abstraction library without any changes to the application source code at all and found that we could. We then investigated how JIT compilation could be integrated into existing HPC applications. Kripke and Laghos were presented here, and we demonstrated that the existing dispatch schemes could be replaced with the proposed JIT-compilation mechanism. This replacement produced significant speedups to the AoT compilation process, which increases programmer productivity, and moreover, the resulting code is simpler and easier to maintain, which also increases programmer productivity. In the future, we expect to see uses of JIT compilation replacing generic algorithm implementations in HPC applications which represent so many {\it potential} specializations that instantiating them during AoT compilation is not practical. We already see this in machine-learning frameworks (e.g., in TensorFlow/XLA) and other domain-specific libraries, and now ClangJIT makes this capability available in the general-purpose C++ programming language. The way that HPC developers think about the construction of optimized kernels, unlike in the past, will increasingly include the availability of JIT compilation capabilities.

\end{section}

\begin{acks}                            
This research was supported by the Exascale Computing Project (\grantnum{GS100000001}{17-SC-20-SC}), a collaborative effort of two \grantsponsor{GS100000001}{U.S. Department of Energy}{} organizations (Office of Science and the National Nuclear Security Administration) responsible for the planning and preparation of a capable exascale ecosystem, including software, applications, hardware, advanced system engineering, and early testbed platforms, in support of the nation's exascale computing imperative. Additionally, this research used resources of the Argonne Leadership Computing Facility, which is a DOE Office of Science User Facility supported under Contract \grantnum{GS100000001}{DE-AC02-06CH11357}. Work at Lawrence Livermore National Laboratory was performed under Contract \grantnum{GS100000001}{DE-AC52-07NA27344} (LLNL-CONF-772305). We would additionally like to thank the many members of the C++ standards committee who provided feedback on this concept during the 2019 committee meeting in Kona. Finally, we thank Quinn Finkel for editing and providing feedback.

\vspace{2em}
\noindent\fbox{\parbox{0.45\textwidth}{The submitted manuscript has been created by UChicago Argonne, LLC, Operator of Argonne National Laboratory ("Argonne"). Argonne, a U.S. Department of Energy Office of Science laboratory, is operated under Contract No. DE-AC02-06CH11357. The U.S. Government retains for itself, and others acting on its behalf, a paid-up nonexclusive, irrevocable worldwide
license in said article to reproduce, prepare derivative works, distribute copies to the public, and perform publicly and display publicly, by or on behalf of the Government. The Department of
Energy will provide public access to these results of federally sponsored research in accordance with the DOE Public Access Plan. \url{http://energy.gov/downloads/doe-public-access-plan}}}
\end{acks}

\bibliography{clangjit}



\end{document}